# FINITE TEMPERATURE PHASE TRANSITION IN QCD WITH STRANGE QUARK: STUDY WITH WILSON FERMIONS ON THE LATTICE*

K. KANAYA
*Institute of Physics, University of Tsukuba, Tsukuba 305, Japan*
E-mail: kanaya@rccp.tsukuba.ac.jp

ABSTRACT

The effect of the strange quark in the finite temperature phase transition of QCD is studied on the lattice. Using the one-plaquette gauge action and the Wilson quark action, the transition in the chiral limit is shown to be continuous for the case of degenerate two flavors, $N_F = 2$, while it is of first order for $N_F \geq 3$. For a more realistic case of massless up and down quarks and a light strange quark, $N_F = 2 + 1$, clear two state signals are observed both for $m_s \simeq 150$ and 400 MeV. In contrast to a previous result with staggered quarks, this suggests a first order transition in the real world. In order to see the implication of these results to the continuum limit, we started to study these issues using improved actions. First results using a RG improved gauge action combined with the standard Wilson quark is presented for the case of $N_F = 2$: With this action the finite temperature transition is shown to be continuous in the chiral limit confirming the result of the standard action. Furthermore, not like the case of the standard action where lattice artifacts make the transition once very strong at intermediate values of the hopping parameter $K$ on $N_t = 4$ and 6 lattices, a smooth crossover is found for the improved action when we increase $1/K - 1/K_c$, in accord with a naive expectation about the fate of second order chiral transition at finite $m_q$.

## 1. Introduction

At sufficiently high temperature, QCD looses two of its characteristic properties: the confinement and the spontaneous breakdown of chiral symmetry. Studies on the lattice have shown that the nature of the corresponding phase transition/crossover sensitively depends on the number of flavors $N_F$ and quark masses: With degenerate $N_F$ quarks, the transition is first order in the chiral limit for $N_F \geq 3$, while a continuous transition is found for $N_F = 2$. At sufficiently large quark mass the transition is smoothed away into a crossover. Therefore the nature of the transition in the real world may depend sensitively on the values of $m_q$'s, especially on the s-quark mass $m_s$ because $m_s$ is of the same order of magnitude as the estimated transition temperature

*In collaboration with Y. Iwasaki, T. Kaneko, S. Kaya, S. Sakai and T. Yoshié.

$T \sim 100 - 200$ MeV,

In this paper, I report our recent study of finite temperature QCD transition on the lattice. We use Wilson fermions as lattice quarks. The investigation using Wilson quarks is highly important because the Wilson formalism of fermions is the only known formalism which possesses a local action for any number of flavors with arbitrary quark masses. We mainly perform simulations on lattices with the temporal extension $N_t = 4$. Although $N_t = 4$ is not large enough to obtain the result for the continuum limit, this work is a first step toward the understanding of the nature of the QCD phase transition with Wilson quarks.

We first study this using the standard one-plaquette action for the gauge part. The $N_F$-dependence of the chiral transition is discussed in the next section.[1] I then discuss the influence of the s-quark for the case of degenerate three quarks ($N_F = 3$) and the case of two massless u and d-qaurks with a light s-quark ($N_F = 2+1$).[2]

In order to see the implication to the continuum limit, a careful estimation of lattice artifacts is required. This motivated us to study these issues using improved actions. In Sect. 3, first results using a RG improved gauge action combined with standard Wilson quark is presented for the case of $N_F = 2$.[3,4]

## 2. Chiral transition[1]

In the coupling parameter space $(\beta, K)$ of lattice QCD, where $\beta = 6/g^2$ is the gauge coupling and $K$ is the hopping parameter, the location of the chiral limit $K_c$ is defined as the point where pion mass $m_\pi$ vanishes in the chirally broken phase. Alternatively we can define $K_c$ as a point where the quark mass $m_q$ vanishes,[5,6] with $m_q$ defined through an axial vector Ward identity.[7,8] The location of the finite temperature transition/crossover $K_t$ can be easily identified numerically by rapid changes of physical observables such as the plaquette, the Polyakov loop and hadron masses. Our study has shown that, for $N_F \leq 6$, the line $K_t$ crosses the line $K_c$ at finite $\beta$.[6,1] Phase diagram for $N_F = 2$ and 3 is given in Fig. 1.

In order to study the nature of the transition at the crossing point $\beta_{ct}$ of $K_c$ and $K_t$, we perform simulations on the line $K_c(\beta)$ starting from a $\beta > \beta_{ct}$ and reduce $\beta$. The number of iterations $N_{\rm inv}$ needed for the quark matrix inversion provides a good indicator to discriminate the deconfining phase from the confining phase, especially on $K_c$, because $N_{\rm inv}$ is enormously large on the $K_c$ line in the confining phase while it is of order several hundreds in the deconfining phase. This difference is due to the appearance of zero modes around $K_c$ in the confining phase[9]. Combining this with measurements of physical observables, we identify the crossing point $\beta_{ct}$. We call this method as "on-$K_c$" simulation method.[1,4] We can check that $\beta_{ct}$ thus determined is consistent with an extrapolation of the $K_t$ line toward the chiral limit (Fig. 1). From the behavior of physical quantities toward $\beta_{ct}$, we are able to determine the order of the chiral transition.

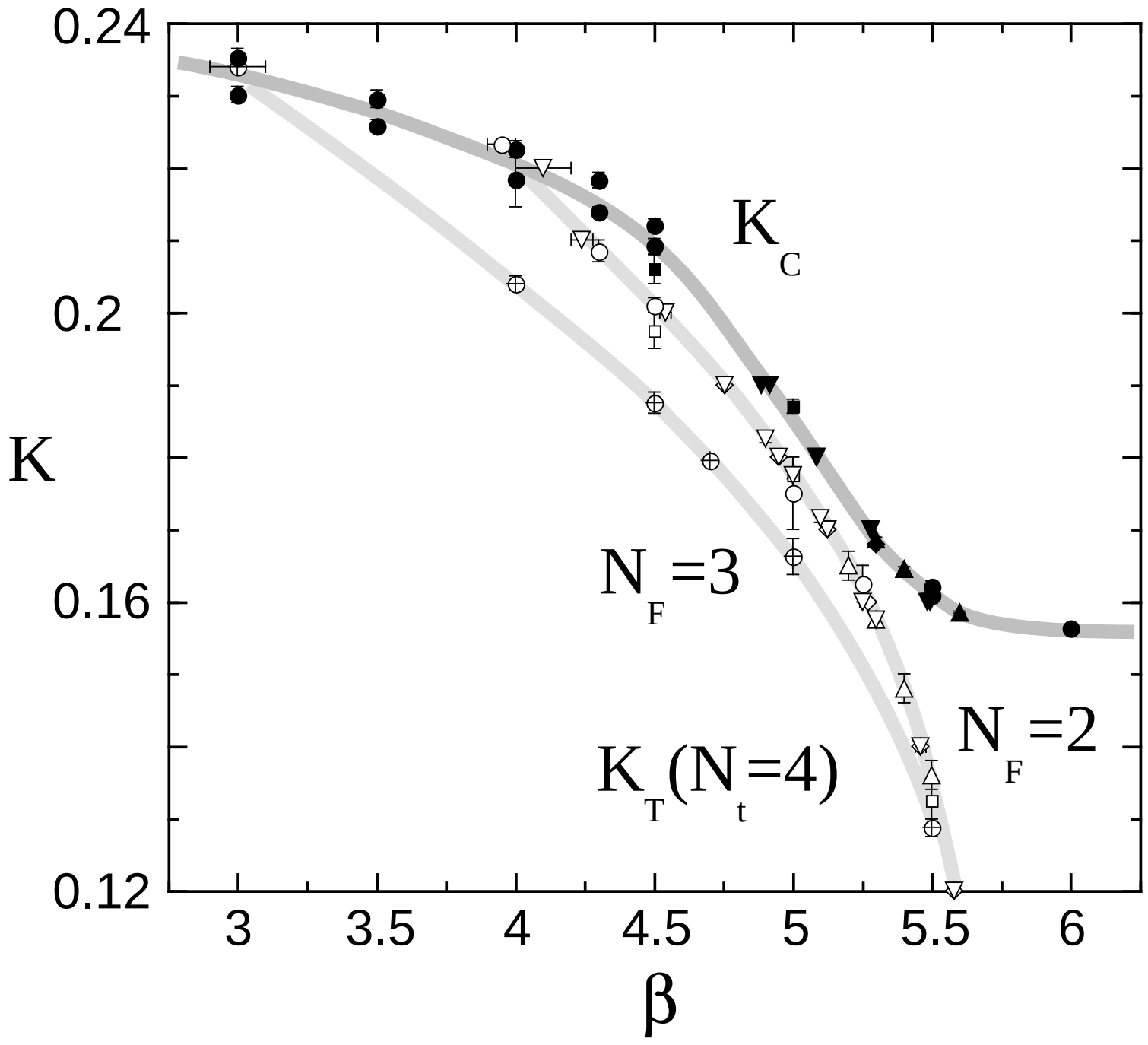

Fig. 1. Phase diagram for $N_F = 2$ and 3. Filled symbols are for $K_c(m_\pi^2)$ and $K_c(m_q)$. Open symbols are for $K_t(N_t = 4)$ for $N_F = 2$ while open circles with cross for $N_F = 3$. Recent results of various groups are collected (see references in 4). Circles (filled, open and those with +) are our data. Shaded lines are to guide the eye.

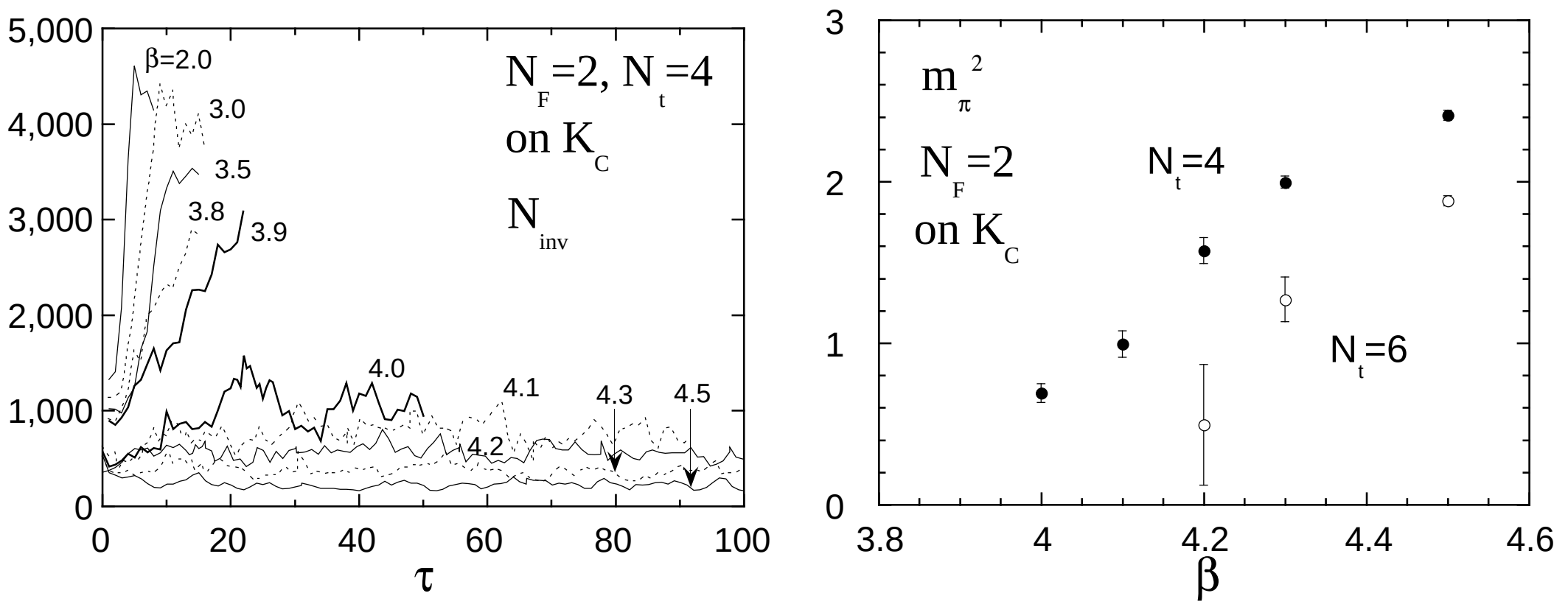

Fig. 2. (a) Time history of $N_{\rm inv}$ and (b) $\beta$-dependence of $(m_\pi a)^2$ for $N_F = 2$ on $K_c$ obtained on an $8^2 \times 10 \times 4$ lattice.

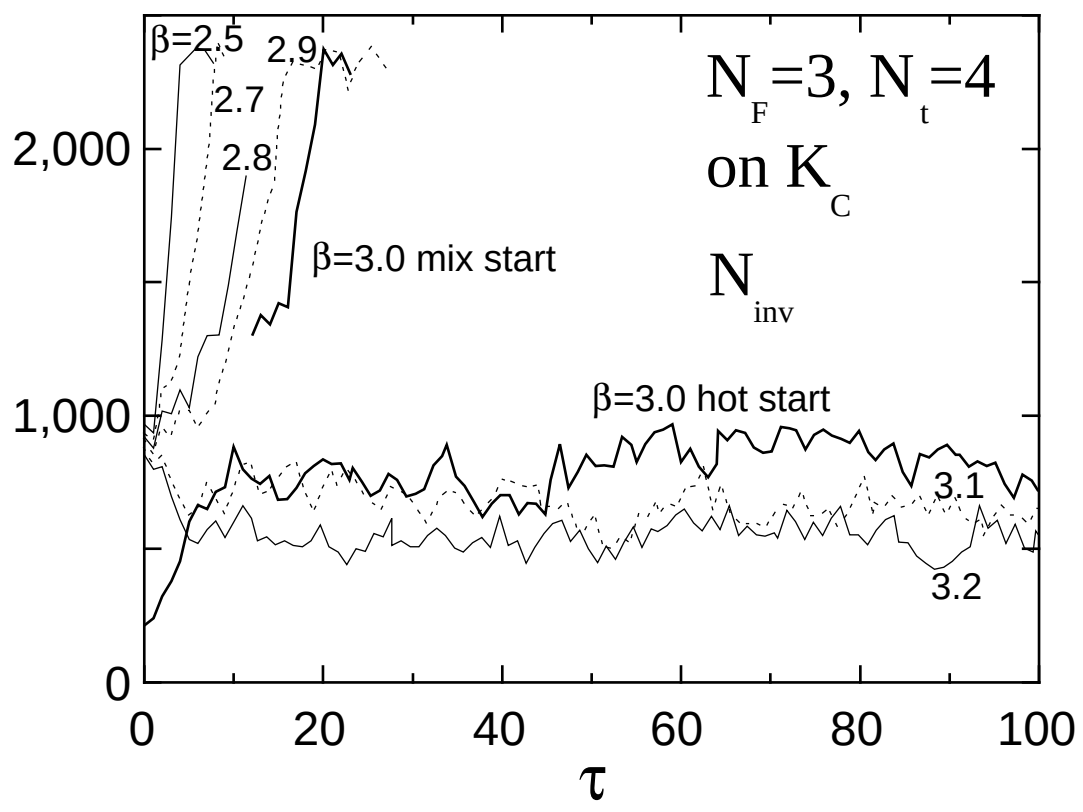

Fig. 3. Time history of $N_{\rm inv}$ for $N_F = 3$ on $K_c$ obtained on an $8^2 \times 10 \times 4$ lattice.

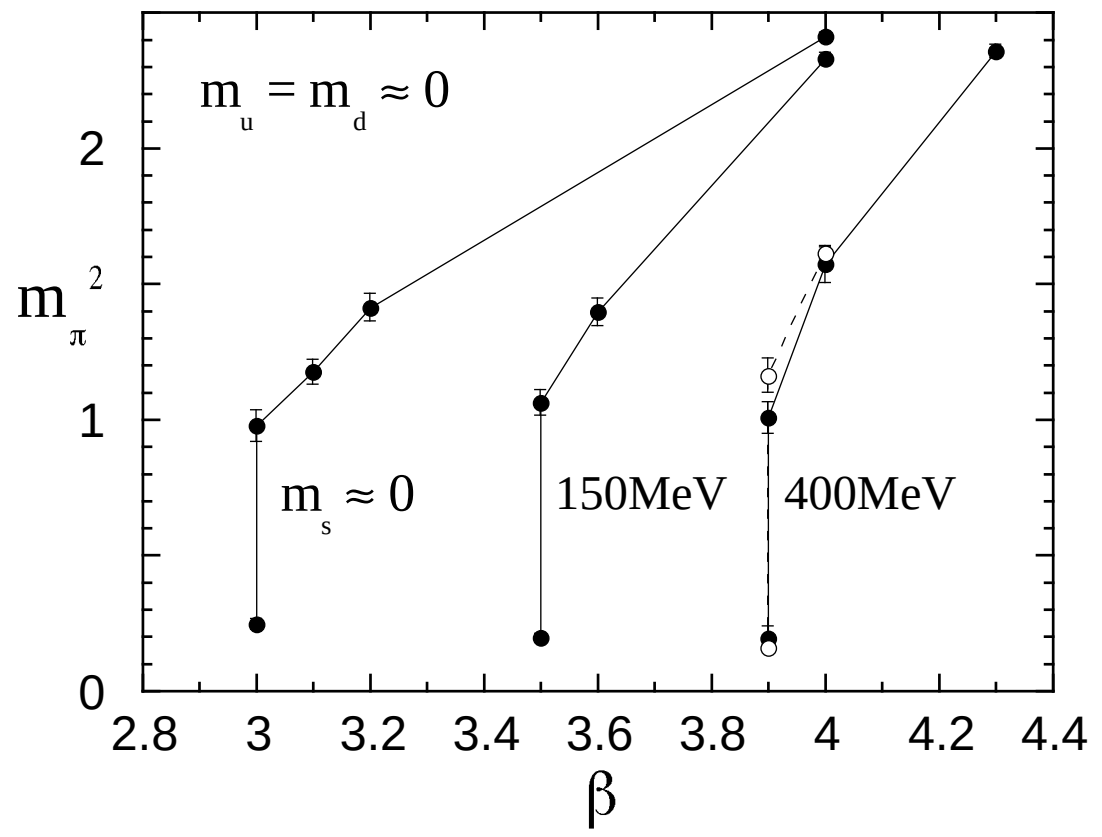

Fig. 4. $(m_\pi a)^2$ for $m_s \simeq 0$, 150 and 400 MeV with $m_{ud} \simeq 0$. Filled and open symbols are for $8^2 \times 10 \times 4$ and $12^3 \times 4$ lattice, respectively.

Time history of $N_{\rm inv}$ for $N_F = 2$ on the $K_c$-line is shown in Fig. 2(a) for $N_t = 4$. We find that when $\beta \geq 4.0$, $N_{\rm inv}$ stays around several hundreds, while $\beta \leq 3.9$ it increases rapidly with $\tau$. Time histories of the plaquette, the Polyakov loop, $m_\pi$ etc. confirm that the system is developing into a confined state for $\beta \leq 3.9$. Therefore we conclude that $\beta_{ct} \sim 3.9 - 4.0$ for $N_t = 4$. Similar study on an $N_t = 6$ lattice gives $\beta_{ct} \sim 4.0 - 4.2$. For both $N_t$, we find no two-state signals around $\beta_{ct}$. $m_\pi^2$ on the $K_c$-line is shown in Fig. 2(b). We find that, when we decrease $\beta$ toward $\beta_{ct}$, $m_\pi^2$ decreases to zero smoothly. This suggests that the transition for $N_F = 2$ is continuous in the chiral limit, in accord with the result of an universality argument[10] and scaling studies with staggered quarks.[11,12]

Corresponding results for $N_F = 3$ are given in Fig. 3 and Fig. 4. (In Fig. 4 $m_s \simeq 0$ corresponds to the case of $N_F = 3$ on $K_c$. Results for $N_F = 2 + 1$ will be discussed in the next section.) In sharp contrast with the case of $N_F = 2$, we here find a clear two-state signal at $\beta = 3.0 \sim \beta_{ct}$: For a hot start, $N_{\rm inv}$ is quite stable around $\sim 800$ and $m_\pi^2$ is large ($\sim 1.0$ in lattice units). On the other hand, for a mix start, $N_{\rm inv}$ shows a rapid increase with $\tau$ and exceeds 2,000 in $\tau \sim 20$, and in accord with this, $m_\pi^2$ the plaquette, the Polyakov loop etc. decreases with $\tau$. This implies a first order chiral transition for $N_F = 3$. These results are again consistent with the prediction based on universality[10].

## 3. The effect of the strange quark[2]

As discussed in the introduction the mass of the s-quark may play a decisive role on the nature of the QCD transition. Let us begin with the case of degenerate $N_F = 3$ and increase $m_q$ by increasing $\beta$ along $K_t$ shown in Fig. 1. On the $K_t$-line we observe clear two-state signals for $\beta \leq 4.7$ (Fig. 5), while no such signals for $\beta \geq 5.0$. At the transition point (in the confining phase) of $\beta = 4.7$ we get $m_q a = 0.175(2)$ and $(m_\pi/m_\rho) = 0.873(6)$. Using our estimation of $a^{-1} \sim 0.8$ GeV for these $\beta$, the critical quark mass $m_q^{crit}$ up to which the first order phase transition persists can therefore be bounded from below as $m_q^{crit} \gtrsim 140$ MeV.[a]

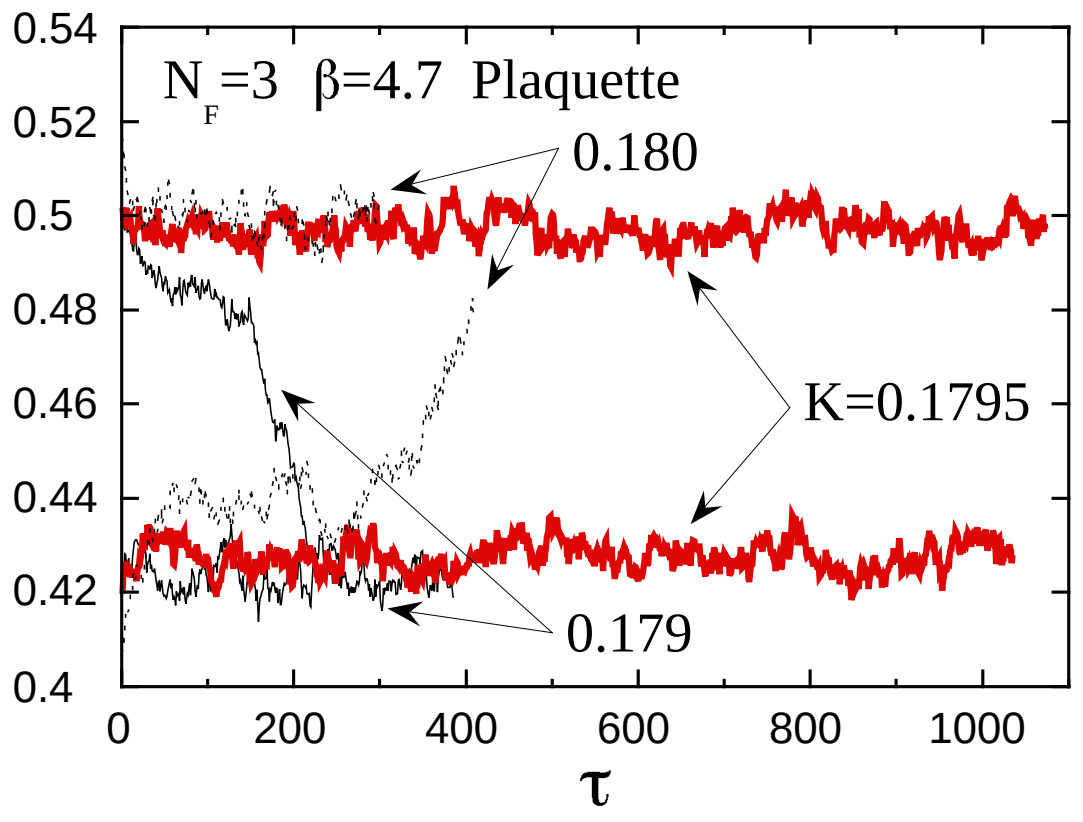

Fig. 5. Time history of the plaquette for $N_F = 3$ at $\beta = 4.7$ obtained on a $12^3 \times 4$ lattice.

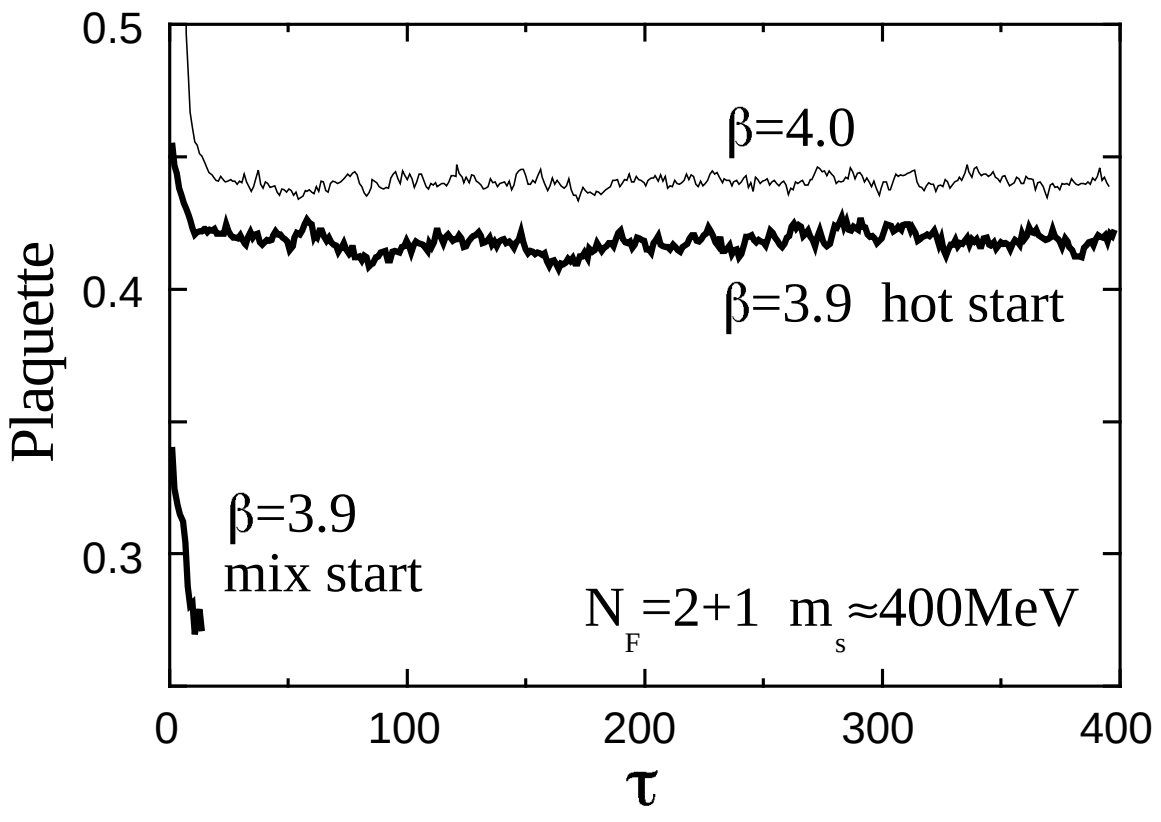

Fig. 6. Time history of the plaquette for $N_F = 2+1$ at $m_{ud} \simeq 0$ and $m_s \simeq 400$ MeV obtained on a $12^3 \times 4$ lattice.

We note that these values are much larger than that obtained previously with staggered quarks[13,14]: $m_q^{crit}a = 0.025 - 0.075$. Using $a^{-1} \sim 0.5$ GeV at $\beta \sim 5.2$ obtained by an extrapolation of the data for $N_F = 2$ [15] we find $m_q^{crit} \sim 10 - 40$ MeV.[b] Using a result of hadron masses at these $\beta$ for $N_F = 4$[16] (because the data for $N_F = 3$ is not available), these $m_q^{crit}$ correspond to $(m_\pi/m_\rho)^{crit} \simeq 0.42 - 0.58$.

We now study the case of $N_F = 2+1$: $m_u = m_d < m_s$. When $m_s$ is reduced from $\infty$ to 0, the phase transition must change from continuous to first order at some quark mass $m_s^*$. In this connection, we note that, when $m_s < m_q^{crit}$ (the critical quark mass for degenerate $N_F = 3$), we expect a first-order transition, while when $m_s \geq m_{ud} > m_q^{crit}$, we expect a crossover.

We study two cases $m_s \simeq 150$ MeV and 400 MeV. We apply the "on-$K_c$" simulation method: Keeping $K_{ud} = K_c$, i.e. $m_{ud} \simeq 0$, and $K_s$ at the point corresponding to the cases $m_s \simeq 150$ or 400 MeV, we decrease $\beta$ until we hit $K_t$. Fig. 6 is an example of time history we get. We find two state signals for both $m_s = 150$ and 400 MeV.[2] $m_\pi^2$ on the $K_c$-line is shown in Fig. 4.

These results imply that $m_s^* \gtrsim 400$ MeV. With staggered quarks Columbia group[14] reported that no transition occurs at $m_{ud}a = 0.025$ and $m_s a = 0.1$ ($m_{ud} \sim 12$ MeV, $m_s \sim 50$ MeV using $a^{-1} \sim 0.5$ GeV). Albeit slightly heavier masses for u and d quarks, this suggests a discrepancy between two lattice quarks. Our results on the nature of the transition for degenerate $N_F = 3$ and $N_F = 2 + 1$ are summarized in Fig. 7.

[a] Our results of the hadron spectrum in the range of $\beta = 3.0 - 4.7$ for $N_F = 2$ and 3 show that physical s-quark mass determined from $m_\phi = 1020$ MeV is about 150 MeV for our definition of $m_q$.
[b] $a^{-1}$ of staggered quarks for $N_F = 3$ may be slightly (about 0.1 GeV) larger. This small difference does not affect our discussions below.

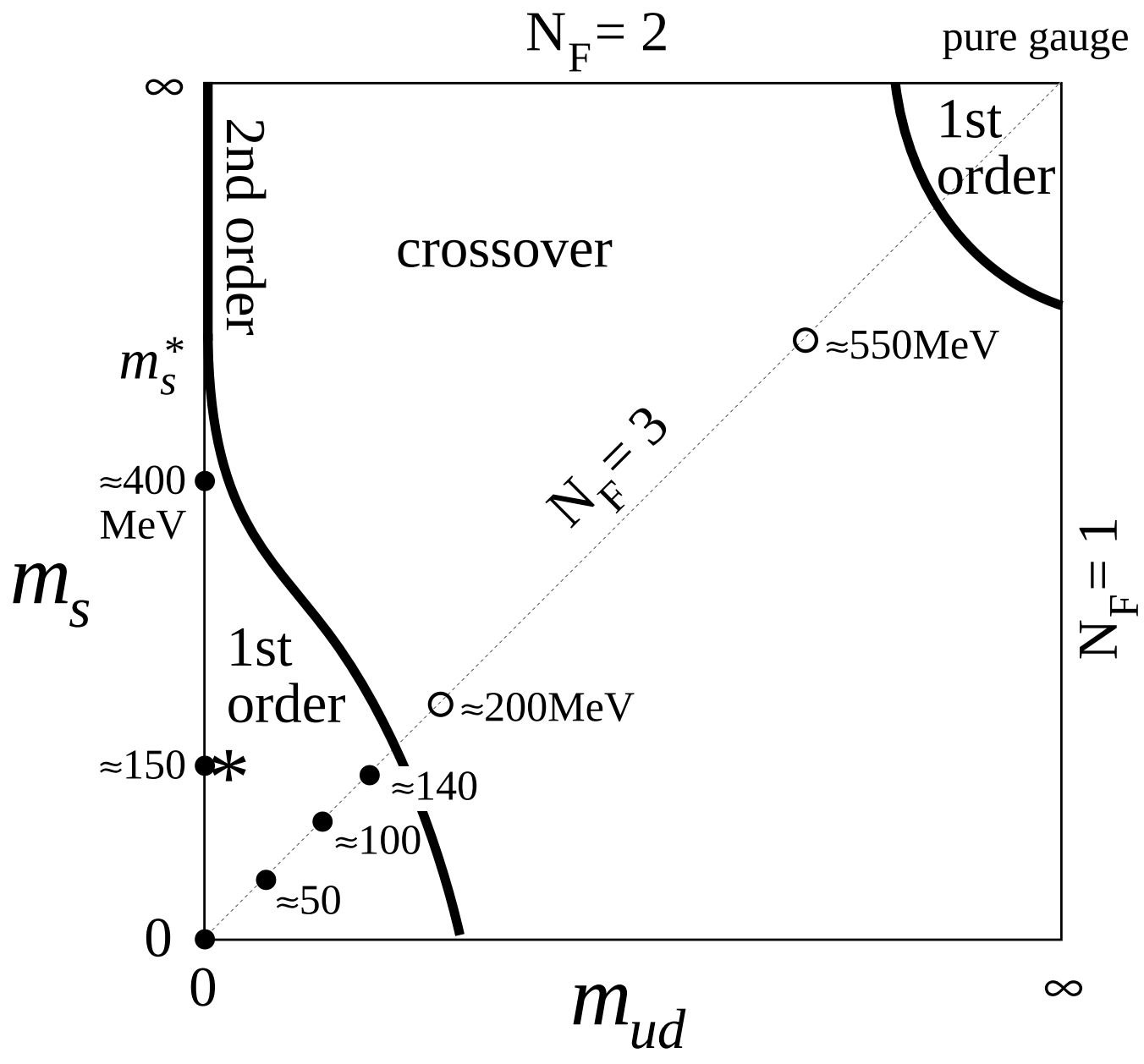

Fig. 7. Order of the finite temperature QCD transition in the $(m_{ud}, m_s)$ plane. First order signals are observed at the points marked with filled circle, while no clear two state signals are found at the points with open circle. The values of quark mass in physical units are computed using $a^{-1} \sim 0.8$ GeV. The real world corresponds to the point marked with star. The second order transition line is suggested to deviate from the vertical axis as $m_{ud} \propto (m_s^* - m_s)^{5/2}$ below $m_s^*$.

## 4. Improved action

In the previous sections we have seen that, although the results of Wilson quarks and staggered quarks are qualitatively consistent, quantitative predictions about the values of critical quark masses are discrepant yet on $N_t = 4$ lattices. We need to get closer to the continuum limit to resolve the discrepancy. This motivated us to study these issues using improved actions.

Based on a study of a block transformation, Iwasaki proposed a RG improved gauge action about ten years ago[17]

$$S_g = 1/g^2\{c_0 \sum(\text{plaquette}) + c_1 \sum(1 \times 2 \text{ loop})\}$$

with $c_1 = -0.331$ and $c_0 = 1 - 8c_1$.[c] As a first step toward an improved full QCD,

[c]This choice of coupling parameters minimizes the distance to the renormalized trajectory after one block transformation. A perturbative calculation shows that the scale parameter of this improved action $\Lambda_{IM}$ is close to that in the $\overline{MS}$ scheme: $\Lambda_{\overline{MS}}/\Lambda_{IM} \simeq 0.488$,[18] in contrast with the large ratio for the case of the standard action. This action was applied to numerical studies of the string tension, the hadron spectrum, topological properties and the $U(1)$ problem[19].

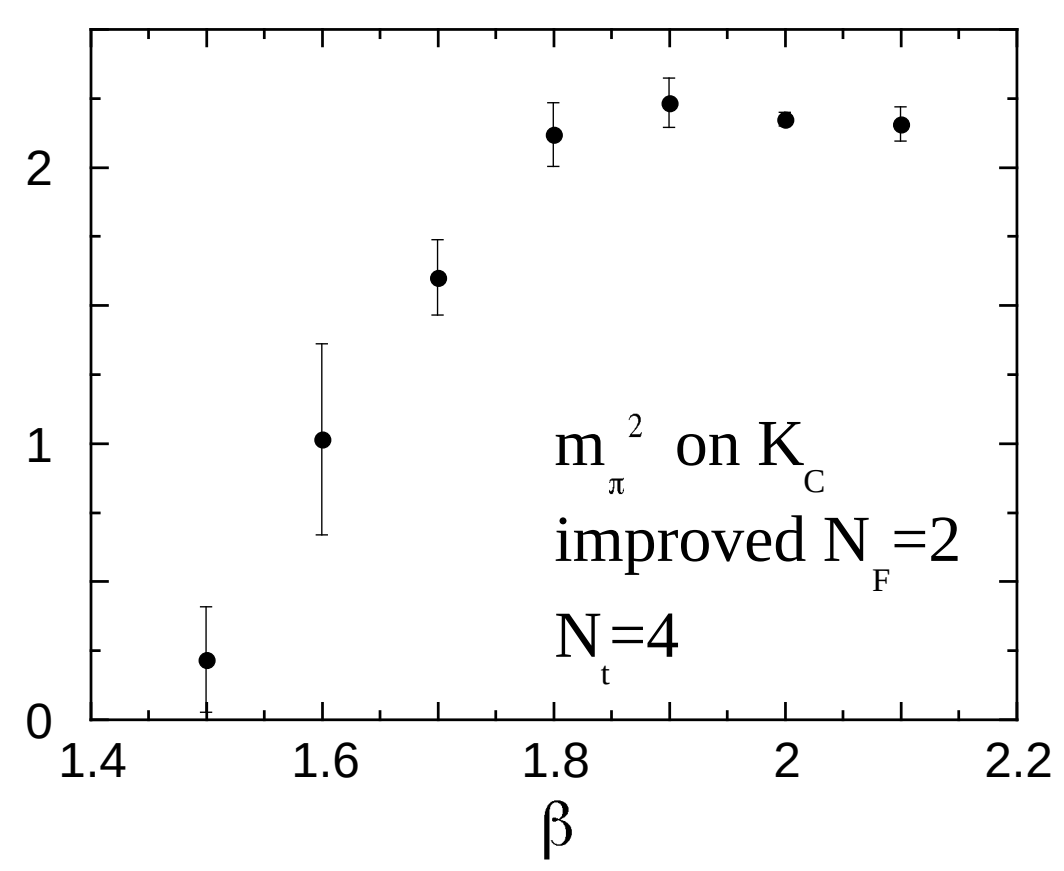

Fig. 8. $(m_\pi a)^2$ from an improved $N_F = 2$ action on $K_c$ obtained on an $8^3 \times 4$ lattice. $K_c$ is determined by the vanishing point of $m_\pi^2$ on an $8^4$ lattice.

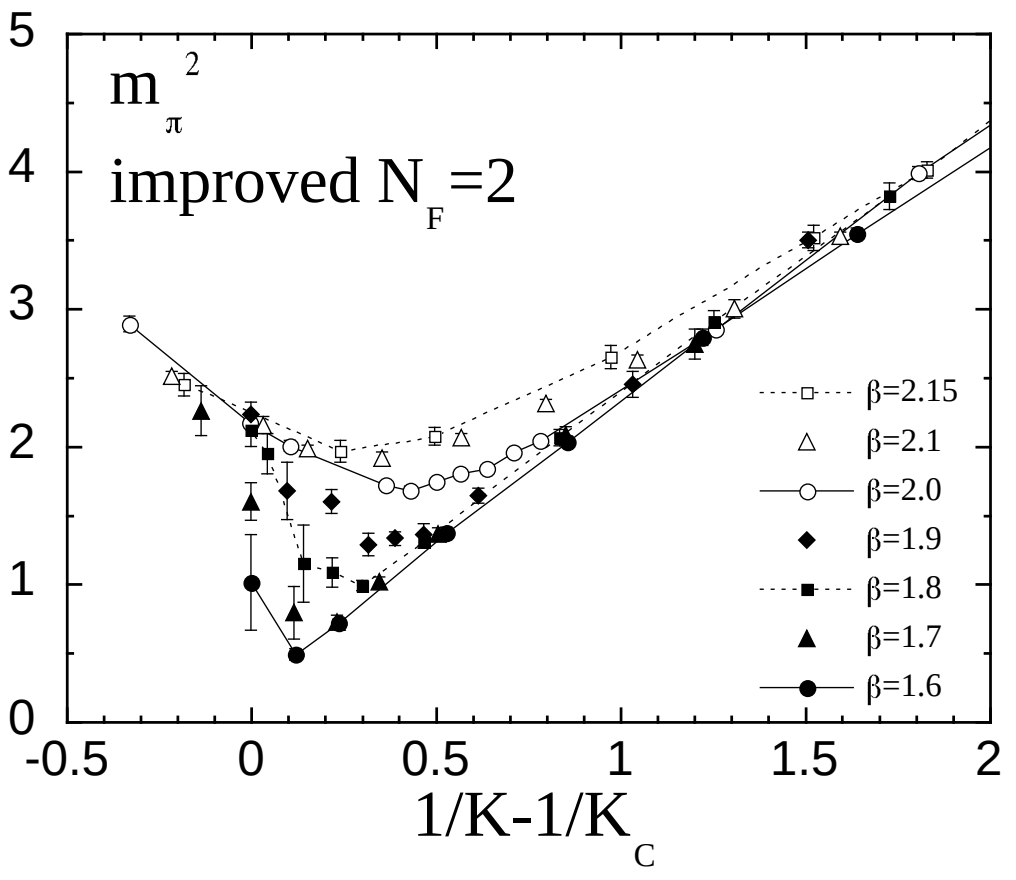

Fig. 9. $(m_\pi a)^2$ from an improved $N_F = 2$ action as a function of $1/K - 1/K_c$ obtained on an $8^3 \times 4$ lattice.

we combine this RG improved pure gauge action with the standard Wilson action.[3,4] This should work well at small $K$ where the effects of quarks can be absorbed by renormalizations of pure gauge interactions. Our numerical results suggests, however, that our action improves the theory very much also near $K_c$.[3,4]

$m_\pi^2$ on $K_c$ is shown in Fig. 8. We see that $m_\pi^2$ decreases smoothly to zero when we decrease $\beta$ toward $\beta_{ct} \sim 1.4 - 1.5$. Therefore, as in the case of the standard action, the finite temperature transition is continuous in the chiral limit. Shown in Fig. 9 is the $1/K - 1/K_c\,(\sim m_q)$ dependence of $m_\pi^2$ for various $\beta$. The straight line envelop in the right corresponds to the chiral behavior $m_\pi^2 \propto m_q$ in the confining phase and deviation from this line signals the transition to the deconfining phase. We find that $m_\pi^2$, as well as other physical observables, is quite smooth at large $\beta$. This strongly suggests that the transition there is actually a crossover, in accord with a naive expectation that the second order chiral transition will become a crossover at finite $m_q$. This result is in clear contrast with that with the standard action where lattice artifacts make the transition once very strong at intermediate $K$ on $N_t = 4$ and 6 lattices.[20]

## 5. Conclusions

We have studied the nature of the finite temperature transition in QCD with Wilson quarks for $N_F = 2$, 3 and 2+1. Using the standard one-plaquette action for the gauge part, the chiral transition is shown to be continuous for $N_F = 2$ on both $N_t = 4$ and $N_t = 6$ lattices. For $N_F = 3$ at $N_t = 4$, clear two state signals are observed on the $K_t$-line for $m_q \lesssim 140$ MeV. For $N_F = 2 + 1$ we have studied the cases $m_s \simeq 150$ and 400 MeV with $m_u = m_d \simeq 0$, and we have found two state signals for both cases. These results suggest a first order QCD transition in the real world. Our results on the nature of the transition for degenerate $N_F = 3$ and $N_F = 2+1$ are summarized in Fig. 7.

We also studied these issues using a RG improved action. As in the case of the standard action, the finite temperature transition is continuous in the chiral limit for $N_F = 2$. On the other hand, not like the case of the standard action where lattice artifacts make the transition for $N_F = 2$ very strong at intermediate $K$, the transition with the improved action becomes quickly smooth when we increase $1/K - 1/K_c$. This suggests that the transition there is a crossover, in accord with a naive expectation about the fate of second order chiral transition at finite $m_q$. Studies for $N_F = 3$ and $2+1$ are in progress.

I am grateful to my collaborators Y. Iwasaki, S. Kaya, T. Kaneko, S. Sakai, and T. Yoshié, with whom the studies reported here are done. Simulations are performed with HITAC S820/80 at KEK and with QCDPAX and VPP-500/30 at the University of Tsukuba. I thank members of KEK and the other members of QCDPAX collaboration for their support. This work is in part supported by the Grant-in-Aid of Ministry of Education, Science and Culture (No.06NP0401).